%% file: tle-casing.tex
\documentclass[
    paper
]{geophysics}

\usepackage[textsize=footnotesize, textwidth=2.8cm]{todonotes}

\usepackage[utf8]{inputenc}
\usepackage{lmodern}
\usepackage[T1]{fontenc}

\usepackage{amssymb, amsmath, amsfonts}
\usepackage{upquote}
\usepackage[strings]{underscore}
\usepackage{siunitx}                
\usepackage{tabularx}
\usepackage{colortbl}
\usepackage{booktabs}
\usepackage{xspace}
\usepackage{flafter}

\usepackage[pdftex, final]{hyperref}
\hypersetup{allcolors=blue, allbordercolors={0 0 .5}, colorlinks=true}

\DeclareGraphicsExtensions{.pdf,.png,.jpg}

\ifthenelse{\boolean{@twoc}}{%
  }{%
  }

\ifthenelse{\boolean{@twoc}}{%
  }{%
  }

\begin{document}

\title{Electrical and electromagnetic responses over steel-cased wells}

\renewcommand{\thefootnote}{\fnsymbol{footnote}}

\ms{}  

\author{%
Lindsey J. Heagy\footnotemark[1],
and %
Douglas W. Oldenburg\footnotemark[1], \\
\footnotemark[1]Department of Earth, Ocean and Atmospheric Sciences, University of British Columbia, Vancouver, British Columbia \\
e-mail: \href{mailto:lheagy@eoas.ubc.ca}{lheagy@eoas.ubc.ca}
}

\footer{}
\lefthead{Heagy \& Oldenburg}
\righthead{}

\maketitle

\begin{abstract}
  Electrical and electromagnetic (EM) methods can be diagnostic geophysical imaging tools for monitoring applications, such as carbon capture and storage or hydraulic fracturing. In these settings, it is common that steel-cased wells and other steel infrastructure are present. Grounded source methods, which use electrodes to inject current into the earth are of interest for casing integrity and monitoring applications. Electrostatic, or direct current (DC) resistivity, experiments form the basis of our understanding of the physics of grounded source experiments in terms of charges, currents, and electric fields. Steel-cased wells are highly conductive and although their presence makes numerical modelling more challenging, they can help targets of interest be detected because they channel charges and currents to depth. Time-domain EM experiments use a time-varying transmitter current. Understanding the EM response requires that we consider both galvanic, or DC currents, as well as image currents that are induced in the subsurface. As compared to DC experiments, the physics of EM is more complex because of the multiple current systems, as well as the need to consider magnetic permeability of steel-cased wells. However, EM experiments have the advantage that they can provide a large data set that is sensitive to a target of interest. Ultimately this will increase the potential for being able to extract information about the target.
\end{abstract}

\section{Introduction}
Electrical conductivity can be a diagnostic physical property for problems involving
carbon sequestration, wastewater injections, geothermal energy, hydraulic fracturing, enhanced oil recovery operations, and monitoring changes within a reservoir with time. Information about the conductivity, and carrying out the imaging, can be achieved by using electrical (DC) or electromagnetic (EM) surveys (e.g. \cite{Hoversten2015, Um2015, Puzyrev2017}). In all of these settings, steel infrastructure, including wells, are typically present. This has been demonstrated to be helpful, as steel is highly conductive and can help channel currents to depths much greater than possible than if a well were not present. To illustrate, we consider a simple model of a conductive cylindrical target at a depth of 900m, as shown in Figure \ref{fig:impact-of-wells}. A target can be conductive if injected fluids are more conductive than reservoir fluids, for example, in some wastewater injections, hydraulic fracturing, or geothermal operations. Depending on reservoir pressures, temperatures, and pore-fluid conductivities, the injection of supercritical CO$_2$ can result in a conductive target \citep{Borner2015}. We simulate a direct current (DC) resistivity experiment (also referred to as Electrical Resistivity Tomography, ERT) using the cylindrical meshes implemented in the open-source SimPEG software \citep{Heagy2019a, Cockett2015}. Electric field data are collected on the surface of the earth along a line perpendicular to the current electrodes and these data are shown in Figure \ref{fig:impact-of-wells-data}. For this example, the target would not be detected if no well is present. With a conductive casing, the difference between the data with, and without, the target is $\sim30$\% of the signal and its impact is clearly visible in the data.

\input{figures/impact-of-wells}

\input{figures/impact-of-wells-data}

The presence of casing helps us detect targets at depth. Using steel infrastructure to help detect deep targets is not a new idea. For example, \cite{Sill1978} performed a DC resistivity experiment in which the casing was treated as a long electrode in order to see if a fault system could be detected. \cite{Rocroi1985} used wells to detect a resistive hydrocarbon reservoir, and there are numerous examples of wells being used as ``long electrodes'' in environmental studies (e.g. \cite{Ramirez1996, Rucker2010}).

Although wells can be beneficial for detecting deep targets, their presence also complicates the analysis of electrical and EM data because they are highly conductive and magnetic and are difficult to incorporate into standard numerical modelling tools because of their geometry. Wells are typically millimeters in thickness and may extend for several kilometers. Some early works studied the ``distortion'' of electrical and EM due to wells \citep{Wait1983, Holladay1984, Johnston1987}. With higher quality data being collected and large-scale computational resources now readily accessible, there is renewed interest in understanding EM data in these settings in order to be able to monitor and delineate targets of interest. In addition to monitoring applications, there is also interest in using electrical and electromagnetic methods to assess the integrity of a well from the surface (e.g. \cite{Beskardes2021, Wilt2020}). In these scenarios, the well itself is the target of interest.

The objective of this paper is to examine and discuss the physics of electrical and EM responses of steel-cased wells. We will focus our attention on grounded source methods: DC resistivity and grounded source electromagnetic methods, where a time-varying current is applied. Grounded source methods are of particular interest for monitoring and well-integrity applications because they take advantage of the effect of the steel-cased well channeling currents in these experiments. The other category of experiments is inductive-source EM methods which use a time-varying current through a loop or coil to generate time-varying magnetic fields. We will not discuss these methods, but there is substantial literature on inductive sources within the context of well-logging (e.g. \cite{Wu1994}), cross-well electromagnetics (e.g. \cite{Uchida1991, Nekut1995}), and large-loop surface to borehole experiments \citep{Augustin1989}, and examples that examine the impact of wells on marine CSEM \citep{Swidinsky2013} and in airborne EM surveys \citep{Kang2020a}.

This paper is organized as follows. We start with a discussion of the DC resistivity experiment. This gives us the fundamental building blocks: currents, charges and electric fields, at the electrostatic limit of Maxwell’s equations. Next, we introduce time-variation of fields and fluxes in a time-domain experiment, and we discuss the additional complexity this introduces into the problem. Finally, we illustrate how magnetic permeability impacts EM experiments and discuss how EM provides improved detectability of conductive and resistive targets as compared to DC experiments . To aid in the exploration of these concepts, we have provided a collection of Jupyter notebooks, available at: \href{https://github.com/simpeg-research/heagy-2021-tle-casing}{https://github.com/simpeg-research/heagy-2021-tle-casing} that reproduce figures from this paper \citep{Heagy2021software}.

\section{DC response: currents, charges and electric fields}

The DC resistivity experiment forms the foundation for understanding electromagnetic responses. In a basic experiment, two electrodes, one positive and one negative, inject current into the subsurface. Charges build up where there are contrasts in electrical conductivity (or resistivity, its inverse). Electric potentials (or potential differences) are measured at the surface. Steel casing is a very strong conductor ($\sim 5\times 10^6$ S/m) as compared to the surrounding geology, which is typically less than 1 S/m, and therefore has a significant influence on the DC response.

Kaufman pioneered work to understand the electrostatic response of a steel-cased well \cite{Kaufman1990, Kaufman1993}, motivated by well logging applications. In \cite{Kaufman1990}, he performs an asymptotic analysis assuming an infinitely long well in a resistive whole space. The source is a point charge in the center of the well. To illustrate, we ran a simulation of a long well in a whole space with a positive point-charge in the center. We use a distant return electrode that is 1km away from the well. The simulation is run using a cylindrically symmetric mesh in SimPEG. The conductivity of the casing is $5\times10^6$ S/m and the surrounding geology is 100 $\Omega$m\footnote{Resistivity, $\rho$, (units $\Omega$m) is the inverse of conductivity, $\sigma$ (S/m)}. The well has an outer diameter of 10cm and wall-thickness of 1cm. The fluid inside the casing has the same resistivity as the background (100$\Omega$m). In Figure \ref{fig:kaufman-dc}, we show: (a) the model, (b) the resultant current density, (c) charges, and (d) electric fields in a region near the source. \cite{Kaufman1990} describes the response in three zones based on their proximity to the source: a near zone, intermediate zone and far zone. In the near zone, the electric field has both radial and vertical components, negative charges are present on the inside of the casing, and positive charges are present on the outside of the casing. The near zone is quite localized and its vertical extent is $\sim10$ borehole radii for typical conductivity values of the surrounding geology. In this example, the borehole radius is 5cm, and we can see that the near zone, where negative charges are present on the inner radial wall of the casing, extends $\sim 0.5m$ vertically. If the electrode is connected to the casing, the near zone is effectively not present \citep{Kaufman1993}.

\input{figures/kaufman-dc}

In the intermediate zone, the currents and electric fields are vertical within the borehole and casing. As such, there is no accumulation of charges along the inner casing wall, as no currents cross it. Charges do, however, accumulate on the outer surface of the casing; these generate radially-directed electric fields and currents, which are often referred to as leakage currents, within the formation. At each depth slice through the casing and borehole in the intermediate zone, the electric field is uniform. Current density is the product of the conductivity and electric field, and due to the high conductivity of the casing, most of the current flows within the casing. The vertical extent of the intermediate zone depends on the resistivity contrast between the casing and the surrounding formation and extends beyond several hundred meters before transitioning to the far zone, where the influence of the casing disappears \citep{Kaufman1990}.

The radially directed fields from the casing, and the length of the intermediate zone, have practical implications in the context of well-logging because they delineate the region in which measurements can be made to acquire information about the formation resistivity outside the well. Within the intermediate zone, fields behave like those due to a transmission line \citep{Kaufman1990}, and multiple authors have adopted modeling strategies that approximate the well and surrounding medium as a transmission line \citep{Kong2009, Aldridge2015}.

The work in \cite{Kaufman1993} extended the analysis to consider finite length wells. They discuss two end-member cases: ``short'' wells and ``long'' wells in which there is a transition between linear and exponential decay of the currents. The factors that influence which regime is more representative are the physical properties of the casing and surrounding geology and the length of the casing. This can be summarized by the ``conduction length'' defined by \cite{Schenkel1991}, which is $\delta = \sqrt{\sigma_c A_c \rho}$, where $\sigma_c A_c$ is the conductance of the casing (product of its conductivity and area), and $\rho$ is the resistivity of the surrounding geology. If $L_c$, the length of the casing, is much smaller than the conduction length ($L_c \ll \delta$), then the well is in the ``short'' regime and the currents decay linearly. The other end member is ``long'' wells in which $L_c \gg \delta$ and currents decay exponentially with distance from the source.

To illustrate, we have run a suite of simulations for wells of increasing length. The source is a ``top-casing source'' where the positive electrode is connected to the top of the casing, and the return electrode is 8km away. The physical properties of the casing and surrounding geology are the same as the previous model, as is the radius and thickness of the casing. Figure \ref{fig:finite-wells} compares the currents for different lengths of wells. In (a), we show the downward-going current within the casing along with the approximations for short and long wells from \cite{Kaufman1993}. With the increasing length of the well, there is a transition between the linear decay behavior and the exponential decay. In (b), we show the ``leak-off'' currents, the radial component of the current leaving the casing, as a function of depth. These are equal to the derivative of the casing currents in (a) with depth and thus transition from a constant value (the derivative of a linear decay) to an exponential decay with increasing length. Note that this character is identical to the distribution of charges along the length of the well. There is an interesting increase in charge near the end of the well in all cases, this is due to the interface conditions where the tangential component of the electric field and normal component of current density must be preserved across those interfaces. At the end of the well, we encounter a ``corner'' where both must be preserved. Note that if the background was more conductive, then the conduction length is shorter, and thus the transition from a linear decay of current to an exponential decay occurs at a shorter well-length.

\input{figures/finite-wells}

Understanding the distribution of currents and charges in a DC experiment has several implications for survey design and for numerical modeling. For survey design, the casing can help deliver current to depth, where a target of interest may be. For long wells, we are still limited by the exponential decay with depth. For a casing integrity experiment, several authors have shown that if the well is completely compromised, the response is nearly the same as if the portion of the well below the flaw was missing \citep{Heagy2019, Wilt2020}. When connecting the source to the top of a well, charges are distributed along the continuous, conductive path. However, if the flaw only compromises part of the circumference of the well, then it is undetectable from the surface as charges can still be distributed along the entire length of the well.

With respect to numerical modeling, it is useful to note that the primary controlling factor on the distribution of currents for a given geologic background is the product of the conductivity and cross-sectional area of the casing. If this is preserved, the casing can be treated as a solid cylinder or prism without compromising the accuracy of the solution \citep{Heagy2019}. This approximation will begin to break down if the size of the prism approximating the casing has a larger area than the true casing. To overcome this, several alternative approaches have been developed, including replacing the casing with a distribution of dipoles \citep{Cuevas2014}, or relatedly, the use of a method-of-moments approach \citep{Tang2015, Orujov2020}. Other modelling approaches include using a resistor network approach \citep{Yang2016}, OcTree meshes to locally refine around the casing \citep{Haber2016}, and the development of hierarchical finite element approach \citep{Weiss2017}, among others. These tools have been used to model infrastructure including horizontal wells and settings with multiple wells.

\section{EM response: time-varying fields and fluxes}

Moving from a DC resistivity experiment, at the electrostatic limit of Maxwell’s equations, to an EM experiment with a source waveform that varies in time, introduces two complicating factors: (1) the response is now comprised of galvanic and inductive effects in the earth and casing, and (2) magnetic permeability now  influences the response because steel has a substantial magnetic permeability (> 50 $\mu_0$ \cite{Wu1994}).

Prior to considering the influence of magnetic permeability on the response, we will first examine the EM response of a conductive well. We consider the response in the time domain as this is arguably more intuitive for understanding the elements contributing to the response, as opposed to the frequency domain, where energy is partitioned into in-phase and out-of-phase components. We use the same setup as previously with a 10cm diameter well in a $100 \Omega$m halfspace. The length of the casing is 1km, and the return electrode is 1km radially distant from the well. To perform the simulation, we use a 3D cylindrical mesh that discretizes the azimuthal direction. This allows us to simulate the return electrode as a point (rather than a disc, as would be the case if a cylindrically symmetric mesh is used).

Figure \ref{fig:tdem-currents-cross-section} shows cross-sections of the currents in the Earth for a time-domain experiment with a conductive casing in a halfspace. At t=0ms, the transmitter waveform is still on and we see the steady-state galvanic currents that are observed in a DC-resistivity experiment. For the EM experiment, let’s first focus our attention on the half-space. After t=0ms, the current through the transmitter is immediately shut off. This creates a time-varying magnetic field, which by Lenz' / Faraday's law induces \emph{image currents} in the Earth which oppose that change. This can be observed at t=0.1ms as the current density that is oriented in the negative x-direction between 0 and 1000m (the same direction as the current in the wire prior to shut-off). As time progresses, both the galvanic and image currents diffuse downwards and outwards. Their interaction can be observed as the circulation of current. Shifting our attention to the casing model, we see that the presence of the casing changes the initial distribution of the steady-state currents at t=0. At later times, we observe the circulation of current as we did in the half-space experiment, but we also see some interesting behavior on the other side of the casing (x < 0m). There is a ``shadow zone'' where no current is visible, this is particularly noticeable in the t=0.1ms and 1ms images between approximately x=-1000m and 0m. To understand why this arises, it is helpful to look at a depth slice, which we show in Figure \ref{fig:tdem-currents-depth-slice}.

\input{figures/tdem-currents-cross-section}

The depth slices in Figure \ref{fig:tdem-currents-depth-slice} are of the current density in the half-space (left), casing model (center), and the difference due to the casing (casing minus half-space; right) 10m below the surface at time t=0.1ms. In the half-space model, we are slicing through the image currents, which are oriented right to left (the same direction as the current in the wire originally). In the center, the ``shadow'' or null in the current density is approximately at $x=-800$m, $y=0$m, for this depth and time. This is a 3D effect that is caused by currents being channeled into the conductive casing. The signal due to the casing (casing response minus the half-space) is much simpler. Due to the symmetry of the casing, we can see that this signal is purely radial. In a casing integrity experiment, or a monitoring study with a vertical well, the radial component of the electric field is most sensitive to the features of interest.

\input{figures/tdem-currents-depth-slice}

If a well is deviated or horizontal, the problem is no longer symmetric, and simulating the expected scenario is required to provide insight as to which fields are most sensitive to the targets of interest. Some numerical approaches have been developed for handling complex geometries with conductive casings (e.g. \cite{Haber2016, Weiss2017}).

Another important factor for the EM survey is the fact that the steel casings have a high magnetic permeability. Our above example simulated a conductive well and assumed the magnetic permeability of the casing is equal to that of free space. For simple geometries, such as a vertical well, we can use cylindrical meshes to simulate the impacts of magnetic permeability. However, simulating 3D geometries when magnetic permeability is considered is much more challenging. Unlike conductivity, where preserving the product of the conductivity and the cross-sectional area of the casing is a sufficient approximation if the mesh is not too coarse, no simple rule for ``upscaling'' magnetic permeability has yet been found. By Maxwell’s equations, a large magnetic permeability causes a concentration of magnetic flux in those permeable targets, and this in turn impacts the resultant electric fields and currents. In the next section, we will illustrate the impacts of magnetic permeability on data that are measured at the surface when we have  a cylindrically symmetric model. How to capture and simulate these effects on large, complex 3D problems is an open avenue of research.

\section{Detecting downhole targets}

The previous section established some of the fundamental concepts for the behaviour of currents in an EM experiment with casing present but we have not yet investigated how this impacts our ability to detect a target at depth. To examine this, we revisit the example we first introduced in Figure \ref{fig:impact-of-wells}. Now, we consider a time-domain EM experiment and measure radial electric field data on the surface. Similar to what we illustrated in Figure \ref{fig:tdem-currents-depth-slice}, because of the symmetry of the problem, it is only the radial component of the electric field that is sensitive to the casing and target. We selected three locations along a line perpendicular to the current wire: 300m, 500m, and 700m away from the well, and we plot the amplitude of the electric field as a function of time after shut-off. The top plots show the simulated data for the scenario with the target (solid) and without (dashed) for: (a) a conductive well and (b) a conductive permeable well. In order to show the DC response, we show a zoomed-in view of the very early times in the thin plots in columns 1 and 3. We do not show a scenario without any casing because the target is not measurable in the data. When considering detectability of a target of interest, there are two aspects to consider: (1) if the signal due to the target (difference between with and without) is above the noise floor, and (2) if that difference is a significant percentage of the response. We show both of these plots in the second and third rows of Figure \ref{fig:impact-of-wells-em-data}, respectively.

First, we examine the plots on the left, for a conductive casing. There is a clear benefit of EM as compared to DC when we consider the secondary signal as a percentage of the baseline response. The electrostatic (DC) response is in the 10\% range. As we proceed to later times, the signal is several hundred percent of the baseline response. Importantly, the signal in the range of several to $\sim$15ms is measurable for the noise floor of $10^{-7}$ V/m that we chose. For 1m dipoles, this corresponds to a 100nV noise floor. This is a slightly more conservative estimate than the 20nV noise floor for the 32-bit ZEN receiver from Zonge Engineering \citep{Weiss2016}. The choice of noise floor will depend on the transmitter current, dipole length, and other factors such as the receiver noise characteristics and noise sources at the field site. Even if the noise floor was increased by an order of magnitude, there would still be a substantial portion of the time-series at each location above a 100\% difference.

\input{figures/impact-of-wells-em-data}

We also simulate the scenario where the well has a magnetic permeability of 100 $\mu_0$. The simulated data are plotted on the right side of Figure \ref{fig:impact-of-wells-em-data}. There are a few notable differences. First, the response due to the well in a half-space (dashed lines) decays much more slowly than if the well was only conductive. Over the entire time range we plot, the radial electric field data are above the chosen $10^{-7}$ V/m noise floor. In the center plot, we see that the difference between the scenario with and without the target decays more slowly, so there is a longer time range above which the signal is measurable. However, the maximum difference as a percentage is smaller than if the well was only conductive; it plateaus at $\sim200\%$ difference, which is still substantial. This example also illustrates the importance of including magnetic permeability in simulations and analysis of EM data. If it were neglected, we would have assumed a drastically different ``baseline'' response. Thus making effective use of EM for analysis and inversion will require further development of strategies to handle (often unknown) magnetic permeability.

Similarly, we could consider a resistive target. Using the same geometry, we replace the target with a 1000$\Omega$m target. The simulated data are shown in Figure \ref{fig:impact-of-wells-em-data-resistive}. There are a couple of important differences we see when working with a resistive target: (a) the amplitude of this signal is smaller as compared to a conductive target and (b) the anomalous signal results in a decreased electric field at the surface at times after shut-off.

\input{figures/impact-of-wells-em-data-resistive}

To understand these differences, we plot the total and anomalous currents for a conductive and a resistive target in Figure \ref{fig:tdem-currents-cross-section-target}. The casing is only conductive. The permeable casing scenarios are included in the notebooks associated with this paper, and they show a similar distribution, but slower decay with time. Currents are channeled to a conductive target, whereas they are diverted around a resistive target. If we consider the DC limit, at t=0ms, then in the case of a conductive target, more currents are channeled to the target. Since these channeled currents exit the conductive target into a more resistive background, the target will have positive charges on all outer boundaries, and the casing will have a net negative secondary charge (see also \cite{Weiss2016, Heagy2019}). As a result, the currents near the surface, which we are sensitive to when measuring the electric field, are pointing towards the casing. This is opposite in direction to the initial DC currents along this line. In the data (Figure \ref{fig:impact-of-wells-em-data}(a)), this causes the electric field values to be reduced at the receivers (the dashed line indicating the background response is above the solid line indicating the response with the target). The instant we shut off the transmitter, the image current is induced and begins to diffuse into the earth. This image current is in the same direction as the current in the wire and points inwards towards the casing. This is the same direction as the anomalous response due to the target. Thus, after shut-off, the amplitude of the electric field is larger than if the target were not there.

For a resistive target, currents are diverted around it. At the DC limit, the anomalous charges along the well are positive, and there are both positive and negative charges on the outer interfaces of the target (see \cite{Heagy2019} for further discussion), but the target has a net negative anomalous charge. Thus the anomalous currents and electric fields point away from the well. At t=0, this results in an increase in the electric field measured at the surface. In Figure \ref{fig:impact-of-wells-em-data-resistive}, this is somewhat difficult to see because the anomalous signal is much smaller in amplitude than when the target is conductive. When the transmitter is shut-off, the induced image currents are in the opposite direction, and therefore the anomalous signal results in a decrease in the amplitude of the measured electric field.

Interestingly, there is very little change in the anomalous currents from t=0 to t=1ms for both the conductive and resistive targets. This is consistent with the constant secondary electric field that is shown in the second row of Figures \ref{fig:impact-of-wells-em-data} and \ref{fig:impact-of-wells-em-data-resistive}. In practice, this may prove to be advantageous from a signal-to-noise perspective because in a time domain experiment, we would be collecting multiple early time measurements that are sensitive to this difference.

\input{figures/tdem-currents-cross-section-target}

Even for this seemingly simple model of a cylindrical block in a halfspace with a steel-cased well, the physics is complex. Beyond thinking about charges in a DC resistivity experiment, time-varying fields and fluxes introduce inductive effects. These effects are advantageous for detecting a target at depth, but they can challenge one’s intuition about electromagnetics. Thus numerical modelling and tools for visualizing charges, fields, and fluxes can be extremely valuable for building an understanding of EM responses in these settings.

\section{Conclusions}

This article focussed on forward simulations of grounded source DC and EM experiments with the goal of understanding the fundamental physics and relative merits of the two types of surveys in settings with steel-cased wells. Numerical simulations are instrumental for understanding physical responses and assessing detectability of targets of interest. The first example illustrated how steel-cased wells can help us detect targets at depth in a DC experiment. We are, by no means, the first to show this effect, and other authors have illustrated that additional steel infrastructure, such as multiple wells, can further amplify signals of interest (e.g. \cite{Yang2019}). The story is more complicated though if a current source is connected to a multilateral well where current is then distributed along multiple different wells \citep{Weiss2017}.

The ability to see to greater depths, shown with DC, also extends to EM. There is, however, more complexity because we now have both galvanic and image currents. Thus, EM methods offer opportunities for increased signal and larger volumes of data that are sensitive to features of interest. We demonstrated this using simulations of conductive and resistive targets at depth. If we define the anomalous signal as the difference between simulations with and without the target of interest, we found that the largest difference in amplitude of the signal occurred at shut-off, but this value is often only a few percent of the primary field. At later times, this percentage could be substantially greater. For the example with a conductive target, the anomalous signal exceeded 100\% of the primary. Although smaller in amplitude than the early-time values, they were still above the noise floor. Similar conclusions apply both to a resistive target and a conductive target. However, the amplitude of response for a resistor is much smaller than that of a conductor.

Understanding the implications of a conductive, permeable casing on EM responses is not trivial, and there are open questions about how to include permeability effects in large 3D simulations. However, the overall effects of the permeability are to reduce the amplitude of the anomalous response due to a target of interest while extending the anomaly out further in time. This may have some practical advantages for target detection.

The observations that an EM experiment can enhance anomalous signals as compared to a DC experiment, and that anomalous signals can be observed over a significant time range illustrate that a time domain EM experiment can provide a large data set that is sensitive to a target of interest. Ultimately this will increase the potential for being able to extract information about the target. Similar comments are likely applicable for frequency domain EM but we have not explicitly investigated that in this paper. The fact that an EM survey has enhanced information about the target leads us to another area with opportunities for future research: solving the inverse problem. Given data, the goal is to estimate a model of the Earth that is consistent with those data. The inversion of EM data is well-studied in many applications, and progress has been made for inverting EM data with a vertical electric dipole in borehole to surface EM \citep{Cuevas2021}, as well as on 1D inversions in settings with steel-cased wells \citep{Tietze2015}. Open questions remain for how to handle steel-cased wells and infrastructure in 3D inversions. Accurate forward simulations are an important component. It will also be important to account for the very large sensitivity along the length of the borehole, and develop strategies for when the physical properties of the borehole are unknown. Using time-lapse data and employing additional strategies to constrain the inverse problem may offer productive trajectories.

\section{Acknowledgment}

The authors are grateful to Dr. Michael Wilt and Dr. Chester Weiss for organizing this special issue and for their feedback on an earlier version of the manuscript. We also thank the one anonymous reviewer who provided constructive feedback and suggested we include an example of a resistive target in our analysis. We are also grateful to members of the SimPEG community for collaborative conversations and their contributions to the open-source ecosystem.

\clearpage
\bibliographystyle{seg}
\bibliography{tle-casing}

\end{document}

%% file: figures/impact-of-wells.tex
\begin{figure}[!htb]
    \begin{center}
    \includegraphics[width=1\textwidth]{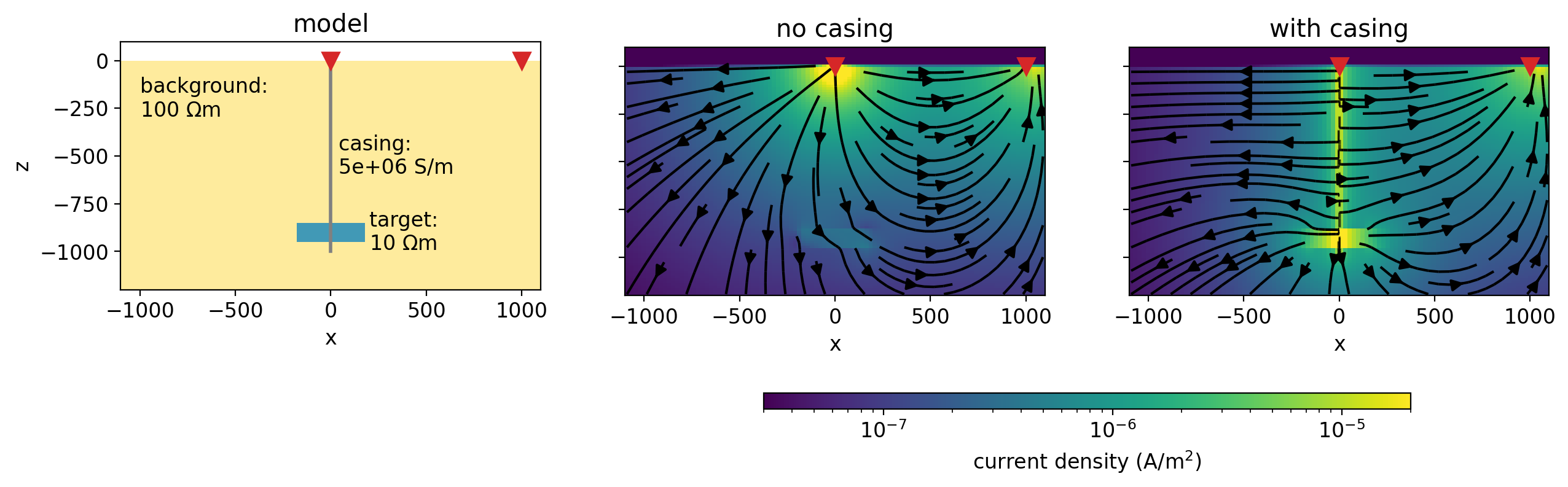}
    \end{center}
\caption{
    Example to illustrate the impact of wells on the ability to detect targets at depth. The image on the left shows the model of a target in a half-space with a steel-cased well. The image in the center shows current density if no casing were present and the image on the right shows the currents with the conductive casing present. The arrows indicate the direction of current flow and the color is the amplitude of the current density.
}
\label{fig:impact-of-wells}
\end{figure}

%% file: figures/impact-of-wells-data.tex
\begin{figure}[!htb]
    \begin{center}
    \includegraphics[width=0.8\textwidth]{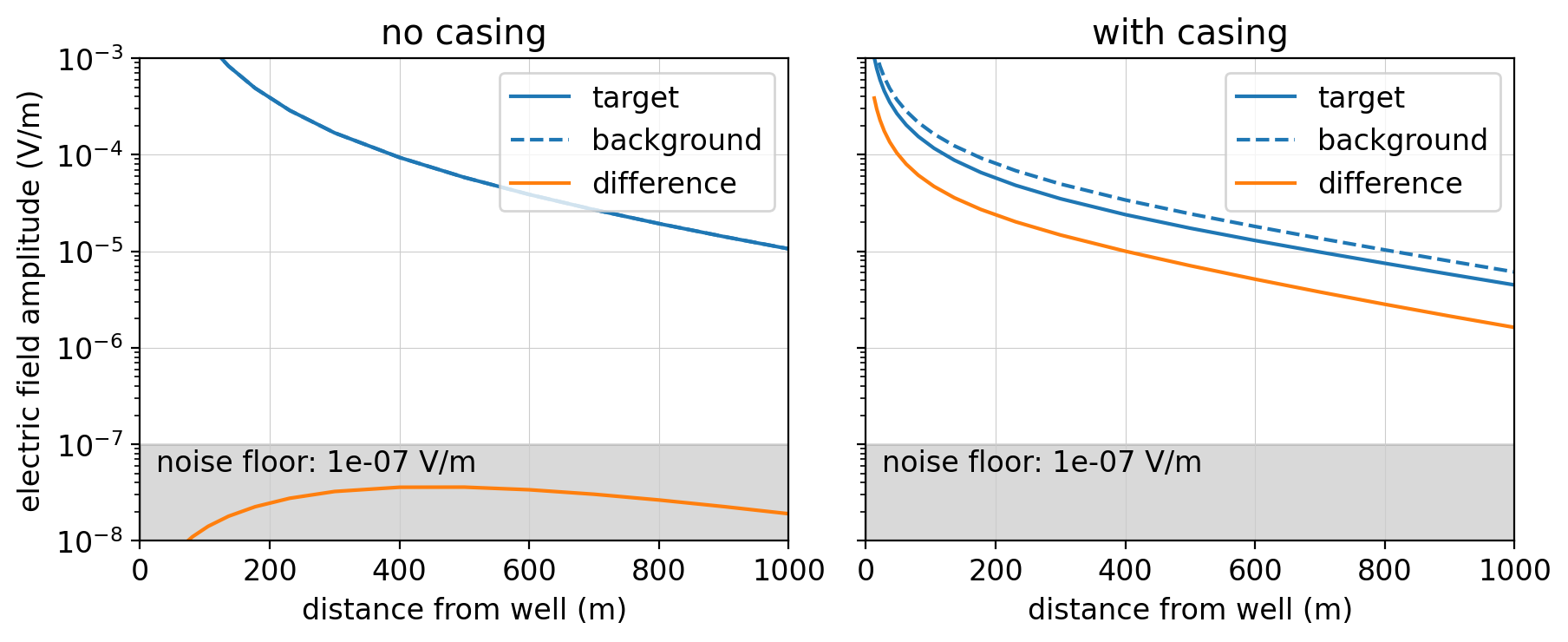}
    \end{center}
\caption{
    Simulated electric field measurements for the DC resistivity experiment shown in Figure \ref{fig:impact-of-wells}.
    The plots show the data with (solid blue) and without (dashed blue) the target. The orange line is the difference between the two; this is the signal due to the target.
    Without the casing, the response due to the target is below a $10^{-7}$ V/m noise floor, whereas with the casing, the signal is detectable.
}
\label{fig:impact-of-wells-data}
\end{figure}

%% file: figures/kaufman-dc.tex
\begin{figure}[!htb]
    \begin{center}
    \includegraphics[width=0.9\textwidth]{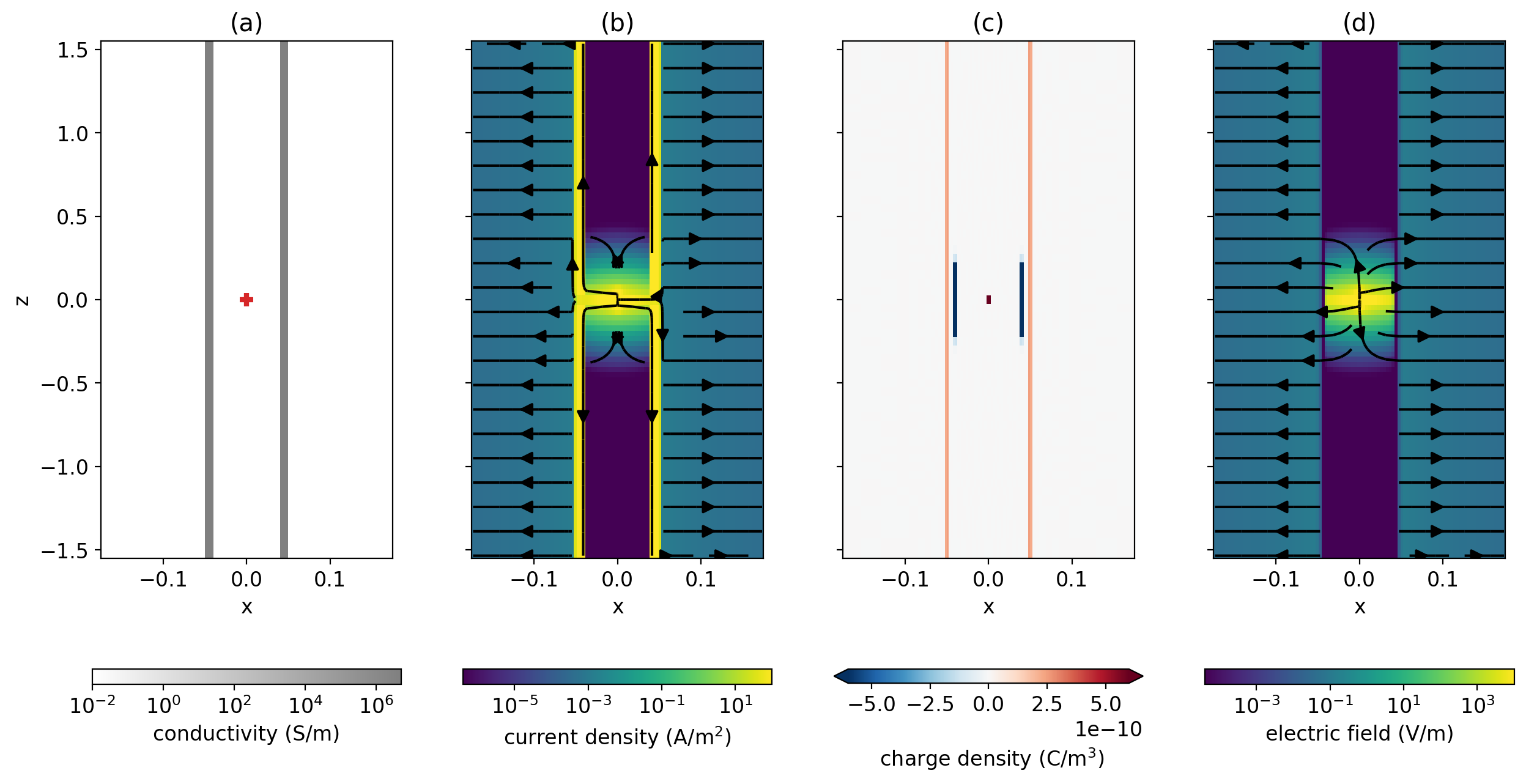}
    \end{center}
\caption{
    DC resistivity experiment where a point source is positioned inside of a long steel-cased well $5\times10^6$ S/m in a $100$ $\Omega$m wholespace. (a) Conductivity model with positive electrode location (red plus); (b) current density; (c) charge density, note that the colorbar has been saturated; (d) electric fields. Figure follows \cite{Heagy2019a}
}
\label{fig:kaufman-dc}
\end{figure}

%% file: figures/finite-wells.tex
\begin{figure}[!htb]
    \begin{center}
    \includegraphics[width=0.55\textwidth]{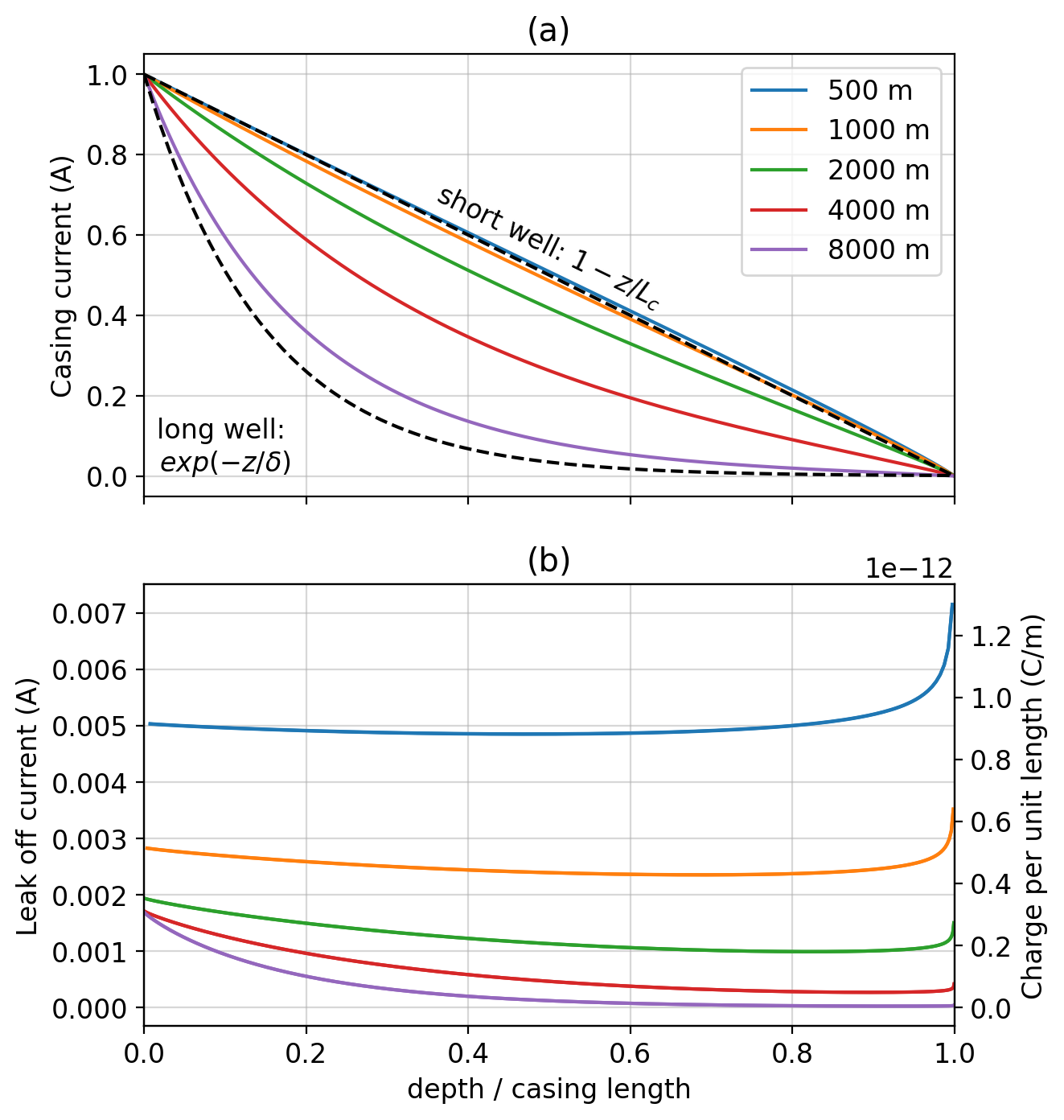}
    \end{center}
\caption{
    Currents in a DC resistivity experiment with the positive electrode connected to the top of the casing.
    (a) Downward-going currents in the casing for different lengths of well. The x-axis is depth normalized by the length of the casing.
    Annotations are the short and long well approximations from \cite{Kaufman1993}. For the long-well approximation, we use $L_c = 8000m$, the length of the longest well included in the simulation.
    (b) Leak-off currents from the well (left axis) and charges on the outer casing wall (right axis).
    Figure follows \cite{Heagy2019a}.
}
\label{fig:finite-wells}
\end{figure}

%% file: figures/tdem-currents-cross-section.tex
\begin{figure}[!htb]
    \begin{center}
    \includegraphics[width=0.8\textwidth]{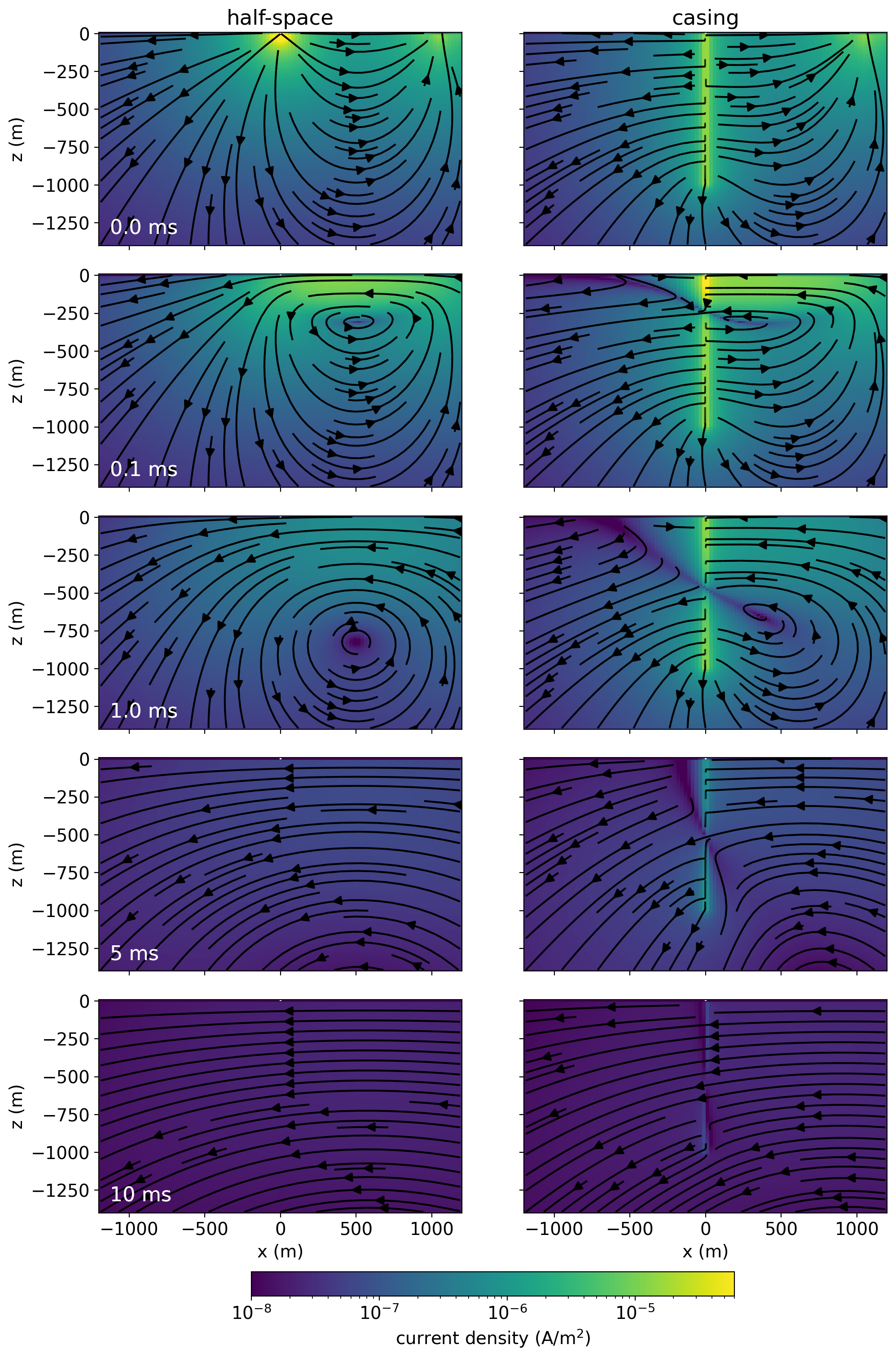}
    \end{center}
\caption{
    Current density for a grounded-source time-domain EM experiment over a $100 \Omega$m half-space (left) and a half-space that includes a 1km steel-cased well (right). The positive electrode is at x=0 and the return electrode is in this cross-section at x=1000m. A step-off waveform is used.
}
\label{fig:tdem-currents-cross-section}
\end{figure}

%% file: figures/tdem-currents-depth-slice.tex
\begin{figure}[!htb]
    \begin{center}
    \includegraphics[width=0.8\textwidth]{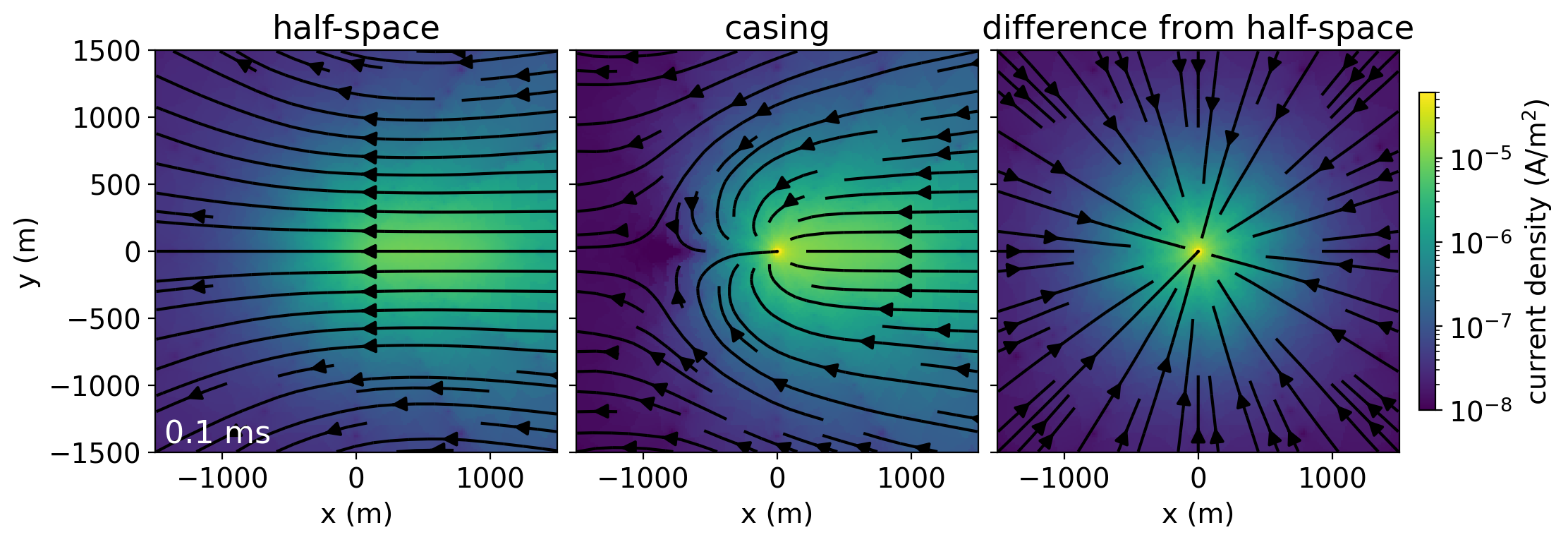}
    \end{center}
\caption{
    Depth slice at z=-10m showing the currents at t=0.1ms for the half-space (left), casing (center) and difference due to the casing (right).
}
\label{fig:tdem-currents-depth-slice}
\end{figure}

%% file: figures/impact-of-wells-em-data.tex
\begin{figure}[!htb]
    \begin{center}
    \includegraphics[width=0.9\textwidth]{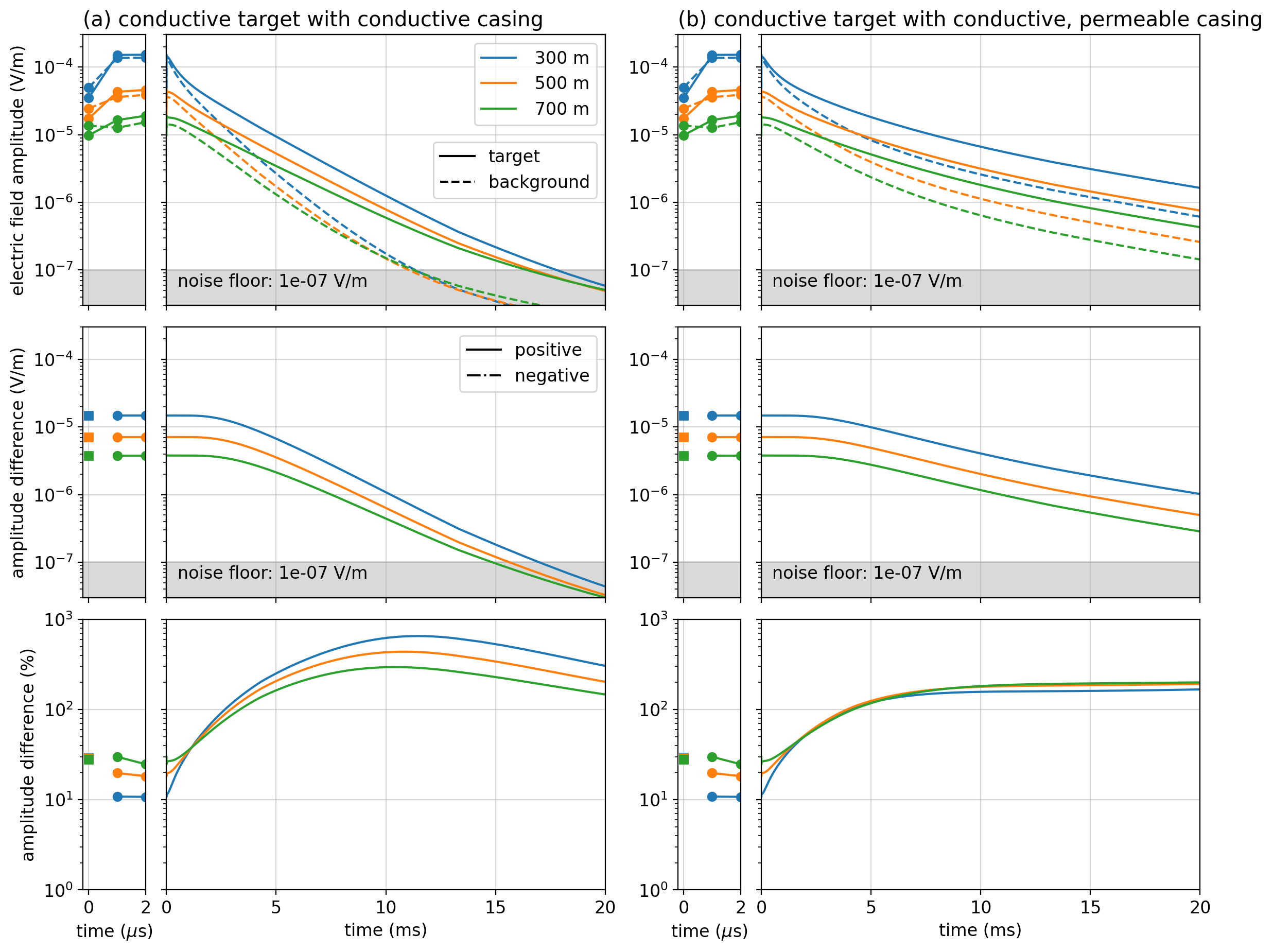}
    \end{center}
\caption{
    Amplitude of radial electric field data in a time-domain EM experiment with a conductive target (10 $\Omega$ m) as shown in Figure \ref{fig:impact-of-wells}.
    The data are collected along a line perpendicular to the transmitter wire, and the color of each line indicates the distance from the well where the timeseries is collected.
    The panels on the left show (a) the simulation for a conductive well which has a magnetic permeability equal to that of free space ($\mu_0$) and on the right, (b) we consider a well that has a permeability of $100 \mu_0$.
    The top plots show the simulated data for the scenario with (solid) and without (dashed) the conductive target. The thin plots on the left zoom in to the earliest times to show the DC response.
    The center plots show the difference between with and without the target. For the earliest times, a circle is used to denote where the amplitude difference is positive (the amplitude with the target is larger than without), and squares are used to show when the difference is negative. The bottom show that difference as a percentage of the results without the target.
}
\label{fig:impact-of-wells-em-data}
\end{figure}

%% file: figures/impact-of-wells-em-data-resistive.tex
\begin{figure}[!htb]
    \begin{center}
    \includegraphics[width=0.9\textwidth]{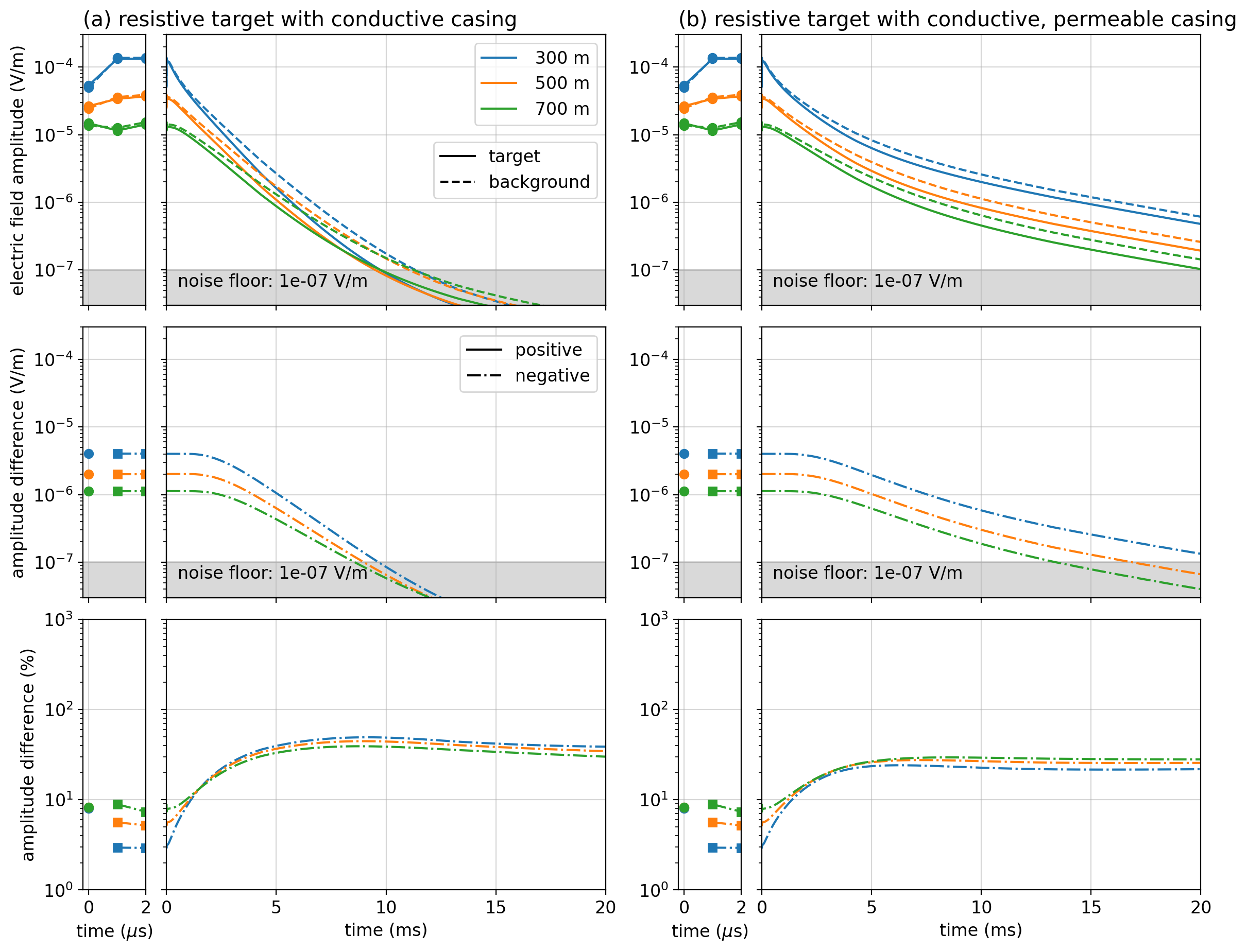}
    \end{center}
\caption{
    Amplitude of the radial electric field data using the same model and survey geometry as Figure \ref{fig:impact-of-wells-em-data} but with a resistive target (1000 $\Omega$ m).
}
\label{fig:impact-of-wells-em-data-resistive}
\end{figure}

%% file: figures/tdem-currents-cross-section-target.tex
\begin{figure}[!htb]
    \begin{center}
    \includegraphics[width=1\textwidth]{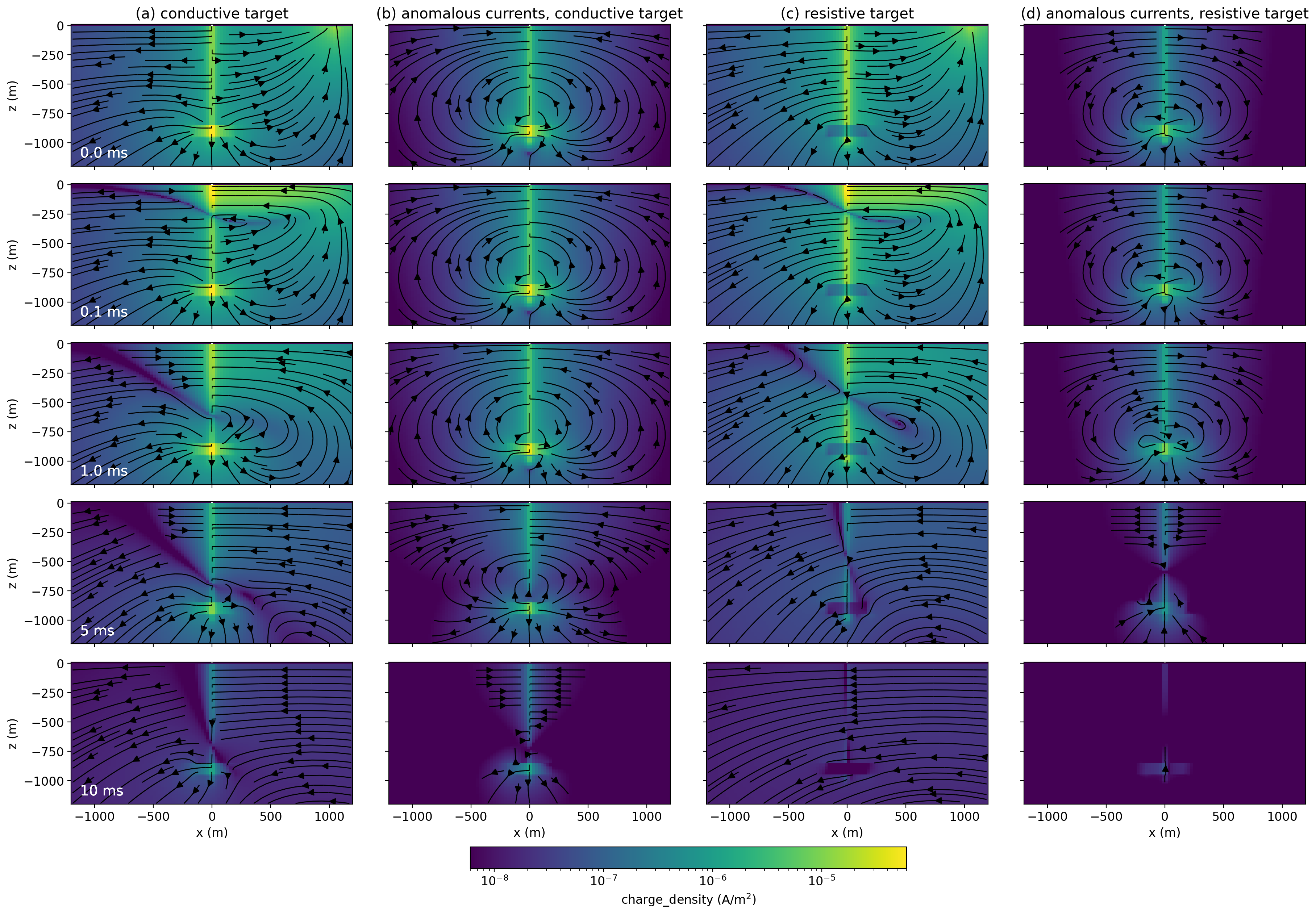}
    \end{center}
\caption{
    (a) Current density for a conductive target (10 $\Omega$m) in a $100 \Omega$m half-space with a purely conductive casing.
    (b) Anomalous current density due to the conductive target (simulation with casing and target minus the simulation of casing in a halfspace).
    (c) Current density for a resistive target (1000 $\Omega$m).
    (d) Anomalous current density due to the resistive target.
}
\label{fig:tdem-currents-cross-section-target}
\end{figure}